# Efficient sunlight promoted nitrogen fixation from air under room temperature and ambient pressure via Ti/Mo composites[1]

Liangchen Chen[a], Jingxuan Shou[b], Yutong Chen[a], Weihang Han[a], Xuewei Tu[a], Luping Zhang[a], Qiang Sun[c,†], Jun Cao[a], Yurong Chang[d], Hui Zheng[a,*]

[a] College of Material, Chemistry and Chemical Engineering, Hangzhou Normal University, Hangzhou 311121, P.R. China

[b] College of civil Engineering, Zhejiang University of Techonlogy, Hangzhou 311121, P.R. China

[c] Australian Research Council Centre of Excellence for Nanoscale BioPhotonics (CNBP), School of Science, RMIT University, Melbourne, VIC 3001, Australia

[d] Henan Qite New Material Co., Ltd., Zhengzhou 450000, P.R. China

† qiang.sun@rmit.edu.au

* huizheng@hznu.edu.cn

**Abstract：**

Photocatalytic nitrogen fixation is an important pathway for carbon neutralization and sustainable development. Inspired by nitrogenase, the participation of molybdenum can effectively activate nitrogen. A novel Ti/Mo composites photocatalyst is designed by sintering the molybdenum acetylacetonate precursor with $TiO_2$. The special carbon-coated hexagonal photocatalyst is obtained which photocatalytic nitrogen fixation performance is enhanced 16 times compared to pure $TiO_2$ at room temperature and ambient pressure. The abundant surface defects in this composite were confirmed to be the key factor for nitrogen fixation. The $^{15}N_2$ isotope labeling experiment was used to demonstrate the feasibility of nitrogen to ammonia conversion. Also, modelling on the interactions between light and the synthesized photocatalyst particle was examined for the light absorption. The optimum nitrogen fixation conditions have been examined, and the nitrogen fixation performance can reach up to 432 $\mu g \cdot g_{cat}^{-1} \cdot h^{-1}$. Numerical simulations via the field-only surface integral method were also carried out to study the interactions between light and the photocatalytic particles to further confirm that it can be a useful material for photocatalyst. This newly developed Ti/Mo composites provide a simple and effective strategy for photocatalytic nitrogen fixation from air directly under ambient conditions.

**Keywords:** Photocatalytic nitrogen fixation; $TiO_2$; Room temperature and normal pressure; sunlight

# 1. Introduction

Ammonia supply is a key element in the economic chain in the world. Ammonia is not only the integral building block of amino acids and nucleotides, but also the crucial precursor in the synthesis of fertilizers, pharmaceuticals, and fibers.[1-3] It is well known that the greenhouse gas of ammonia industry is a major concern for sustainable development. Though 78% nitrogen in air is a handy source of raw material for ammonia industry, the big challenge is to activate nitrogen from its stable N≡N triple bonds (941 kJ/mol).[4, 5] At present, the large-scale industrial nitrogen fixation is usually implemented via Haber-Bosch process to use nitrogen and hydrogen to synthesize ammonia under iron-based catalysts at high temperature and high pressure.[6] This process requires a large amount of fossil energy and emits a significant amount of carbon dioxide.[7-9] Therefore, nitrogen fixation in a green way at room temperature and normal pressure has become one of the important research subjects in chemistry and chemical engineering.

In recent years, photo- and electro- catalysis have been regarded as promising methods of nitrogen fixation and have been widely studied due to their mild reaction conditions at room temperature and normal pressure.[10-15] In particular, photocatalytic nitrogen fixation is considered a more environment-friendly and eco-friendly method and more in line with sustainable development. Photocatalyst materials for nitrogen fixation have been studied including metal oxides, metal sulfides, g-$C_3N_4$, metal-organic frameworks (MOFs) and so on.[16-18] In the study of photocatalytic nitrogen fixation mechanism, an important aspect of semiconductor materials is the oxygen vacancies or other similar defect sites in the materials, and the signal peaks of oxygen vacancies are often characterized by ESR. For photocatalytic nitrogen fixation, the oxygen vacancies on the surface of the material are the active centers for enriching electrons and capturing nitrogen. The N≡N triple bond accepts the electrons provided by the oxygen vacancies and combines with the protons in the solution to form N-H bonds, which are finally converted into ammonia.[19, 20] For example, He *et al.* reported that indium sulfide nanotubes with more sulfur vacancies were prepared by calcining indium sulfide nanotubes to improve the photocatalytic nitrogen fixation effect.[21] Li and his group reported that the oxygen vacancy content increased and the photogenerated carrier separation and nitrogen adsorption capabilities of bismuth molybdate nanosheets were improved by etching bismuth molybdate with sodium hydroxide.[12]

$TiO_2$ is widely used in the field of photocatalysis, due to its stable chemical properties, photochemical stability, and electrons with strong reduction energy generated under illumination, which is often used for photodegradation of pollutants and water decomposition.[22] Anatase is a common crystal form of $TiO_2$ and is an indirect semiconductor. The advantage of this structure is that the generated photoelectrons are not easy to recombine with holes and to annihilate. So it has bright prospects in the field of photocatalytic nitrogen fixation.[23-25] While due to the lack of nitrogen active sites in $TiO_2$, modification methods such as doping and construction of heterojunctions are generally required to improve the application of $TiO_2$ in photocatalytic nitrogen fixation.[26-28] Inspired by the structure of nitrogenase in nature, Mo element plays a key role in Fe-Mo protease. Fe-Mo nitrogenase contains Fe-Mo clusters to form bimetallic cofactors, in which Mo plays a key role.[1] Mo-based materials have been verified by theoretical calculations that have the ability to activate the triple bond of N≡N. Hexagonal molybdenum carbide ($Mo_2C$) is widely used in the field of photoelectric catalysis due to its good electrical conductivity and hydrogen evolution ability.[29-33] Therefore, in this work, to utilizing various virtues of $Mo_2C$ and $TiO_2$ as photocatalysts, a composite material by using $Mo_2C$ and $TiO_2$ for photocatalytic nitrogen fixation was designed and synthesized. Compared with pure $TiO_2$, the Ti/Mo photocatalyst can effectively separate the photogenerated electrons and holes, so that this photocatalyst can convert nitrogen into ammonia efficiently from air directly.

## 2. Experiment

### 2.1 Chemicals and materials

Molybdenum acetylacetonate and glucose were purchased from Adamas-beta Reagent Co., Ltd. Nano-$TiO_2$ and sodium chloride (NaCl) were purchased from Shanghai Hushi Laboratorial Equipment Co., Ltd. N,N-dimethylformamide and absolute ethanol were purchased from Sinopharm Chemical Reagent Co., Ltd. All the purchased materials have been used as analytical reagents without further purification. Deionized water was used in all the experiments.

**2.2 Synthesis of Catalyst**

**2.2.1 Synthesis of Mo$_2$C@C**

Mo$_2$C@C was prepared based on the literature with minor modifications.[31] 0.4 g molybdenum acetylacetonate and 0.14 g glucose were added in a mixture of ethanol/deionized water (5 mL, 4:1). After stirred at 70 ℃ for 15 minutes, 0.1 g of NaCl was added and stirred for another 10 minutes. The solution was then put into an alumina combustion boat to evaporate. The dried powder was put in a tube furnace with ventilate argon for 30 minutes to remove air. After that, the temperature of the furnace was increased to 800 ℃ with the rate of 5 ℃/min, and kept at 800 ℃ for 2 hours before it naturally cooled down to room temperature. The burned black powders were washed with deionized water several times, followed by centrifugation, and dried in a vacuum oven at 90 ℃ to obtain a black solid Mo$_2$C@C. For simplicity, Mo$_2$C@C is named as MC in this work.

**2.2.2 Synthesis of Ti/Mo**

The grinded MC powder was mixed with nano-TiO$_2$ particles with different mass ratios (10 mg, 1:9, 2:8, 3:7, and 4:6) which was added in 10 mL absolute ethanol. Subsequently, the mixture was oscillated by an ultrasonic dispersion for 30 minutes and evaporated to obtain the solid powders. The products are denoted as TMC1, TMC2, TMC3, TMC4, corresponding to the MC mass ratio in the solid powders (10 %, 20 %, 30% and 40 %), respectively.

## 2.3 Characterization of the photocatalyst

The crystal phase of the sample was tested by X-ray diffractometer (XRD, Bruker AXS D8,). The transmission electron microscopy (TEM, FEI Tecnai G2 F20,), the scanning electron microscopy (SEM, Zeiss Sigma500) and energy-dispersive spectroscopy (EDS) were used to analyze the lattice spacing, microscopic morphology and element composition of the sample. The Raman spectrum of the sample was measured by the Raman spectrometer (Senterra II). The elemental composition of the sample was characterized by X-ray photoelectron spectroscopy (XPS, Thermo Scientific K-Alpha). The absorption spectrum was characterized by the UV-visible spectrophotometer (Hitachi U-3900). The fluorescence intensity was characterized by the fluorescence spectrophotometer, and electron spin resonance (ESR) signals were recorded with a Bruker A200 spectrometer.

The photoelectrochemical test (PEC) was performed on the CHI660E electrochemical workstation (Shanghai Chenhua, China) with a standard three-electrode configuration. The Pt plate was used as the logarithmic electrode, and Ag/AgCl was used as the reference electrode. The electrolyte was an aqueous solution of 0.5 M sodium sulfate. The photocurrent response along with time, the *I-t* curve, was measured when a 300W xenon lamp (PLS-SXE300, Beijing Perfect Lamp Co., Ltd.) was used as the light source for intermittent lighting (20 s).

## 2.4 Photocatalytic nitrogen fixation test

The photocatalytic nitrogen fixation experiment was carried out in air at room temperature and normal pressure. Firstly, 5 mg of the catalyst was put into 30 mL of deionized water in the reactor. Then, the mixtures were continuously stirred (700 r/min) in the dark and bubbled with air (80 mL/min) for 30 min, before being exposed to light irradiation. 1 mL initial liquid was taken as the sample to reference. During the reaction, the reactor was irradiated with a 300W xenon lamp (PLS-SXE300, Beijing Perfect Lamp Co., Ltd.), and 1 mL of the reaction solution was collected every hour. The collected solution was filtered with a 0.22 μm filter to remove the catalyst and was detected by the indophenol blue spectrophotometry at 697.5 nm to determine the concentration of $NH_3$ (as $NH_4^+$).

### 2.5 $^{15}N_2$ isotope labelling experimentst

$^{15}N_2$ was used instead of air for photocatalytic nitrogen fixation experiments. The reaction solution was filtered with a 0.2μm filter head, added sulfuric acid to evaporate and concentrated, and added deuterated DMSO-d6 solvent for nuclear magnetic resonance test.

## 3. Results and discussion

### 3.1 Morphology and structure

In order to explore the possible crystal forms of molybdenum carbide, the samples were characterized by XRD. Figure1 shows the XRD patterns of MC, TMC1, TMC2, TMC3 and TMC4. It can be seen that the broad diffraction peaks of MC at 39.4°, 34.4°, and 61.5°correspond to $Mo_2C$ (PDF No. 35-0787). The TMC1, TMC2, TMC3 and TMC4 samples show a broad diffraction peak at 25.3°corresponding to anatase $TiO_2$ (PDF No. 21-1272), and a peak at 27.4°corresponding to rutile $TiO_2$ (PDF No. 21-1276). With the increase of MC, the diffraction peak of $Mo_2C$ at 39.4° becomes more obvious, and the diffraction peak of $TiO_2$ is weakened. The XRD patterns indicate that $TiO_2$ and MC were effectively combined to obtain Ti/Mo photocatalyst by ultrasonic method.

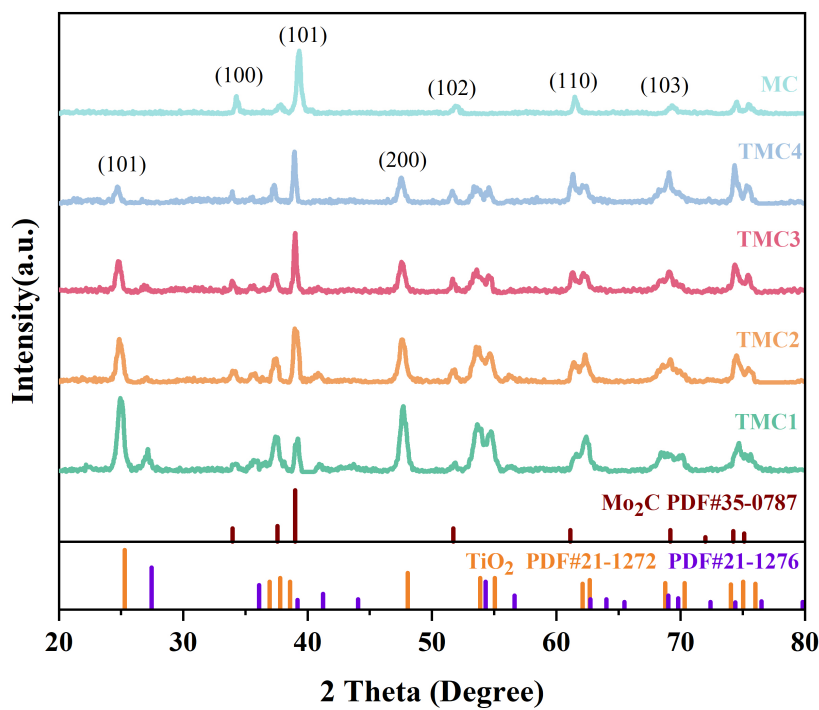

Figure. 1. XRD patterns of pure MC, TMC1, TMC2, TMC3, TMC4 samples.

More characterization of MC and TMC3 were performed, and Figure 2a are the HRTEM images of TMC3. $TiO_2$ can also be observed adhering to the surface in the dark. At the same time, it can be found that the interplanar lattice spacing of 0.352 nm and 0.175 nm corresponding to the (101) plane of $Mo_2C$ and the (101) plane of $TiO_2$ respectively.[34, 35] In the selected area electronic diffraction (SAED) image (Figure 2b), the diffraction rings (103) and (101) correspond to the $TiO_2$ crystal planes, and (102) (103) (104) correspond to the $Mo_2C$ crystal planes. These crystal planes are consistent with the XRD results, which proves the successful synthesis of TMC3.

The micromorphological features of the samples were observed by SEM. Figure 2c shows the SEM image of MC at a scale of 2μm. It is observed that the pure MC has a hexagonal structure, which is consistent with previous literature.[36] Figure 2d shows the image of TMC3 after ultrasonic composite treatment. It is inferred that $TiO_2$ is attached to the surface of MC. Also, the elemental mapping images of TMC3 in Figure 2 (e including $e_1$ to $e_5$) demonstrate almost overlapping distribution of Ti, O, Mo and C elements.

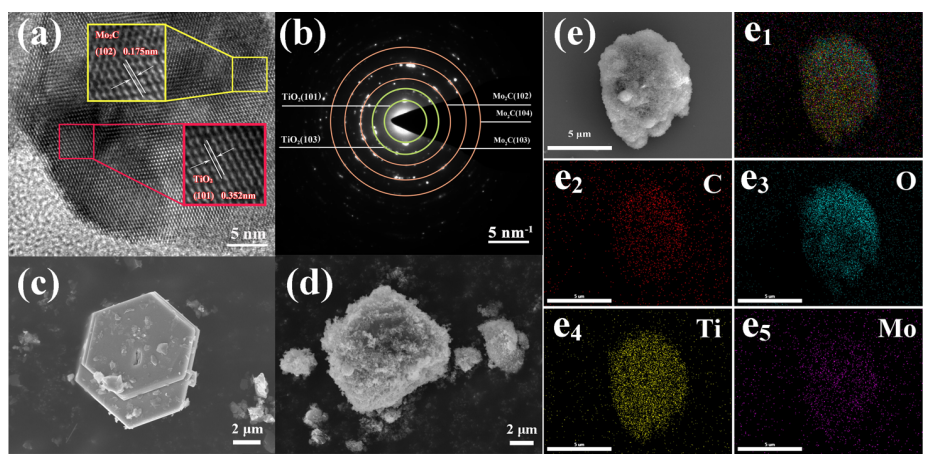

Figure. 2(a) HRTEM image of TMC3; (b) SAED image of TMC3; (c,d) SEM images of MC, TMC3; (e) the elemental mapping image of Ti, O, Mo, C in the TMC3.

To confirm the carbon skeleton in the TMCx samples, the samples were subjected to Raman spectroscopy. The Raman spectrum of the sample is shown in Figure 3a. The D and G bands of carbon at 1333 cm$^{-1}$ and 1600 cm$^{-1}$ in MC without TiO$_2$ indicate that glucose is converted to a carbon skeleton during the preparation process, [36, 37] and the Raman spectrum does not show the characteristic peaks of Mo$_2$C at 228.9 cm$^{-1}$ and 232 cm$^{-1}$. The carbon peak covers the peak of Mo$_2$C so that the characteristic peak of Mo$_2$C cannot be displayed, indicating that the sample has a high graphitization. Figure 3a clearly exhibits four characteristic peaks of TiO$_2$ in TMC1 and TMC2 at 147 cm$^{-1}$, 200 cm$^{-1}$, 448 cm$^{-1}$, 400 cm$^{-1}$, 519 cm$^{-1}$, and 640 cm$^{-1}$. Meanwhile, the characteristic peaks of TiO$_2$ become stronger and the carbon peaks become less significant with the increase of TiO$_2$, which means that TiO$_2$ can be well compounded on the surface of Mo$_2$C@C.

The chemical state and coordination environment of surface elements were investigated by XPS spectroscopy. It can be seen from Figure 3b, most of the elements in MC are carbon and molybdenum. A small amount of oxygen may come from molybdenum oxide due to the oxidation of the sample. It is confirmed again that TMC1 and TMC2 contain four elements Ti, Mo, O, and C, which is consistent with the result of the EDS spectrum. The high-resolution spectrum of the sample TMC3 is shown in Figure 3c-f. In Figure 3c, apart from C-C at 284.7 eV and C=C at 285.4 eV, C-Mo at 283.1 eV can also be found, which is consistent with the literature. [38] Figure 3d shows the Ti-O and O-H peaks of $TiO_2$ at 529.7 eV and 530.3 eV, respectively. The Ti 2p XPS spectrum in Figure 3e shows two main peaks at 458.4 eV and 464.3 eV which attribute to Ti $2p_{3/2}$ and Ti $2p_{1/2}$, respectively. Alos, it can be seen that a small amount of $Ti^{3+}$ exists, which is caused by the crystal lattice defects of $TiO_2$ itself.[39] From the XPS spectrum of Mo 3d (Figure 3f), the binding energy of 227.8 eV and 230.9 eV are attributable to $Mo^{2+}$, 228.4 eV and 232.3 eV to $Mo^{4+}$, and 231.7 eV and 235.2 eV to $Mo^{6+}$.[40, 41] Divalent molybdenum exists in the form of $Mo_2C$, and the presence of tetravalent and hexavalent molybdenum may be caused by the oxidation of the sample.

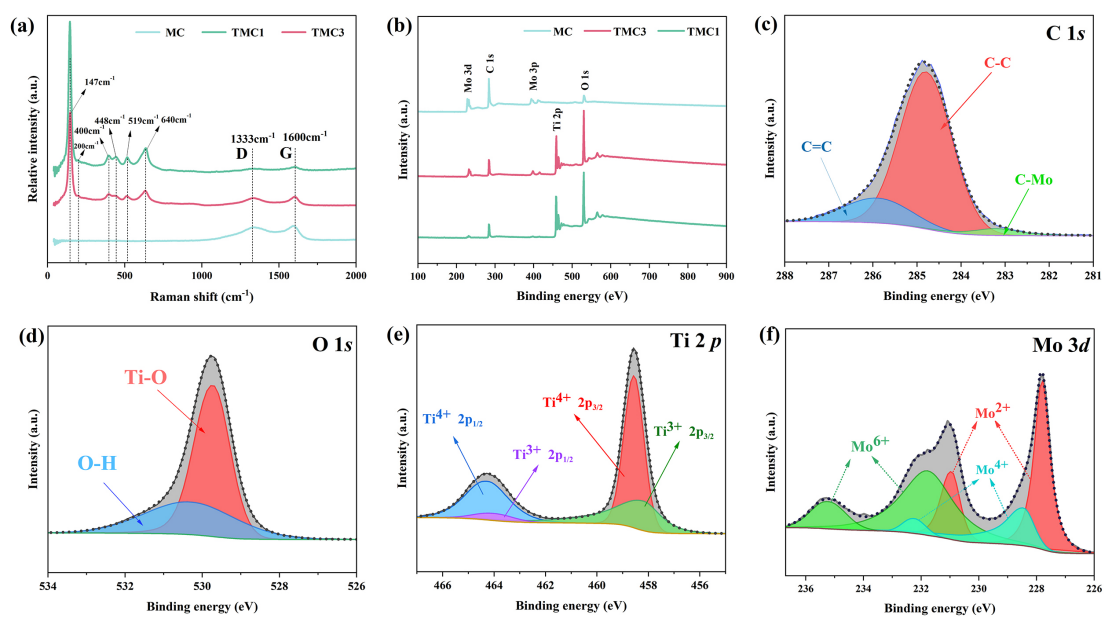

Figure 3. (a) Raman spectra of various MC, TMC1, TMC3; (b) XPS full scan spectra of MC, TMC1, TMC3; (c-f) High-resolution O 1s, C 1s, Ti 2p, Mo 3d spectra of TMC3.

## 3.2 Optical Properties and Surface Features

The absorption performance of all the samples was characterized by using ultraviolet-visible diffuse reflectance spectroscopy (UV-vis). As shown in Figure 4a, the absorption peak of pure $TiO_2$ is at 400 nm, and MC samples loaded with different proportions show absorption bands similar to $TiO_2$. It shows that MC as a co-catalyst support will not affect the absorption band of $TiO_2$. According to previous reports,[42, 43] the Kubelka-Munk function can reflect the change of band gap energy ($E_g$), which can be calculated by the following equation:

$$\alpha h\nu = k(h\nu - E_g)^{n/2} \tag{1}$$

In the above equation, α is the absorption coefficient, h is the Planck constant, ν is the optical frequency, k is a constant, and the band gap energy is expressed by $E_g$. Since $TiO_2$ is an indirect semiconductor which band gap n is 4. As shown in the Figure 4a, the calculated band gap energy ($E_g$) of the sample TMCx (x = 1, 2, 3, 4) and $TiO_2$ are 3.01, 3.04, 3.03, 2.98, 2.98 eV, respectively.[44, 45]

Photoluminescence (PL) spectra can be employed to determine the recombination rate of photogenerated electron-hole pairs. In general, the stronger the fluorescence, the higher the recombination rate of photogenerated electrons and holes in the sample. As shown in Figure 4b, all samples fluoresce at 724 nm. The highest fluorescence intensity is produced by pure $TiO_2$. The fluorescence effect is obviously weakened when MC is loaded. Sample TMC3 exhibits the weakest fluorescence, indicating that this sample has the lowest recombination rate of photogenerated electrons and holes.

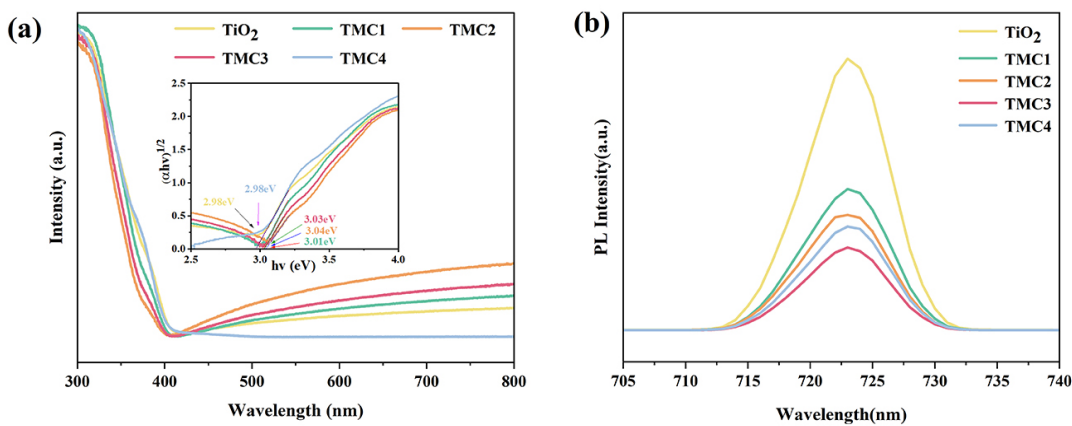

Figure 4. UV–vis absorption spectra profiles; (b) PL spectra of pure $TiO_2$, TMC1, TMC2, TMC3, and TMC4.

To study the adsorption behavior of sample TMC3 to nitrogen, the sample was subjected to nitrogen adsorption analysis and specific surface area test. As shown in Figure 5a, the adsorbed nitrogen volume of pure $TiO_2$ is higher than TMC3, indicating TMC3 has a larger specific surface area. The specific surface area of pure $TiO_2$ and TMC3 was obtained by the BET method, notably the specific area of TMC3 is 562.53 $m^2/g$, which is much higher than that of pure $TiO_2$ (53.48 $m^2/g$). Therefore, $Mo_2C@C$ has a larger specific surface area, which can absorb more $N_2$ and facilitate the reaction. Figure 5b shows the electron paramagnetic resonance (ESR) test diagram of sample TMC3 and $TiO_2$. Sample TMC3 has an obvious peak at g = 2.001 [46-48], and pure $TiO_2$ has no obvious signal peak. This may be due to the carbon vacancy defects caused by the deletion of carbon atoms during the formation of $Mo_2C$ by argon high-temperature sintering. And the lack of some carbon atoms provides active sites for the conversion of nitrogen to ammonia.[49, 50] The defect signal peak is almost undetectable for pure $TiO_2$, while TMC3 has an obvious signal peak. This may be due to the lattice defects caused by high-temperature sintering of inert gases in accordance with the result in Figure 5b, since it is known that the generation of lattice defects provides active sites for the conversion of nitrogen to ammonia.

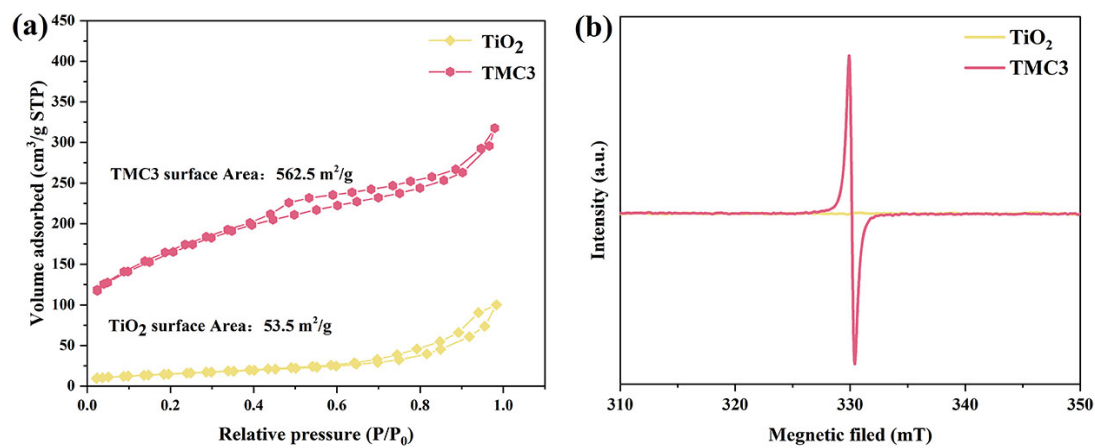

Figure 5. (a) BET adsorption-desorption isotherms of pure TiO$_2$ and TMC3; (b) ESR spectra of pure TiO$_2$ and TMC3.

The EIS measurements were performed on pure TiO$_2$ and TMCx (x=1,2,3,4) to measure the internal resistance for the charge transfer process. As shown in Figure 6a, the sample TMC3 impedance radius is the smallest among all samples. The smaller the radius is, the easier the electron transmission can happen. The loaded TiO$_2$ on MC can transmit photo-generated electrons to Mo$_2$C and increase the transmission rate of photo-generated electrons. At the same time, the photo-generated current process of the samples of pure TiO$_2$ and TMC3 was studied. The transient photocurrent response can directly reflect the life-time of photo-generated charge carriers and the separation efficiency of electron-hole pairs when the xenon lamp is repeatedly switched on and off for 20 seconds. As shown in Figure 6b, the TMC3 photocurrent density of 1.5 μA/cm$^2$ is higher than the pure TiO$_2$ photocurrent density of 1.0 μA/cm$^2$, indicating that the sample TMC3 has a higher photogenerated charge carrier density. The decrease in impedance and the increase in photocurrent are beneficial to converting nitrogen to ammonia.

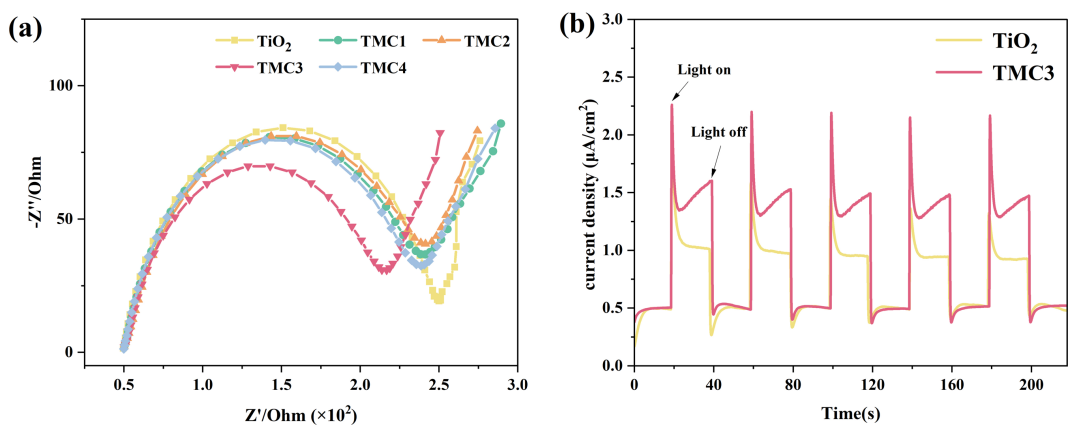

Figure 6. (a) EIS Nyquist plots of pure TiO$_2$ and TMCx(x=1,2,3,4); (b) Transient photocurrent responses of pure TiO$_2$ and TMC3.

### 3.3 Photocatalytic nitrogen fixation performance and mechanism

### 3.3.1 Nitrogen fixation performance of TMCx photocatalyst

In the photocatalytic nitrogen fixation experiment, the catalyst was irradiated by ultraviolet light in an aqueous solution under high-purity nitrogen atmosphere to evaluate the photocatalytic nitrogen fixation performance of all samples. As shown in Figure 7a, the time growth rate of NH$^{4+}$ is relatively stable, and the content of NH$^{4+}$ continues to increase with time. The growth rate of NH$^{4+}$ of TMC3 sample is about 432 μg·g$_{cat}^{-1}$·h$^{-1}$, which is 16 times that of pure TiO$_2$. The ammonium radical detection method adopts indophenol blue spectrophotometry to detect at 697.5 nm. Sample TMC3 was subjected to simulated sunlight-catalytic nitrogen fixation experiments under a xenon lamp (300 w) with a solar filter (Figure 7b). The ammonia production efficiency was about 353 μg·g$_{cat}^{-1}$·h$^{-1}$. When the TMC3 sample was test in Ar atmosphere or without light (Figure 7b), it can be seen that the generation of ammonium needs to be carried out under light with nitrogen as the required raw material. The stability performance was also examined. The apparent quantum efficiency (AQE = 0.1%) of sample TMC3 was obtained by the nitrogen fixation experiment with a single-band 365 nm light source, the calculation formula based on literature [1] is as follows:

$$\text{AQE (\%)} = \frac{N_e}{N_p} \times 100\% = \frac{3 \times \text{The number of molecules of NH}_3 \text{ produced}}{\text{The number of incident photons}} \times 100\% \quad (2)$$

It can be seen from Figure 7c that the catalytic performance of the catalyst remains basically stable in 5 cycles per 5 h, and the XRD patterns (Figure S3) basically unchanged after the reaction, indicating that the sample is stable. The $^{15}$N isotope $N_2$ labelling method proved the reliability of the conversion of $N_2$ to ammonia (Figure S4). Substituting $^{15}N_2$ for $^{14}N_2$ in photocatalytic nitrogen fixation reaction, then acidifying the reaction solution, $^{15}NH_4^+$ was detected by nuclear magnetism. This result shows that the nitrogen source comes from the external nitrogen atmosphere directly.

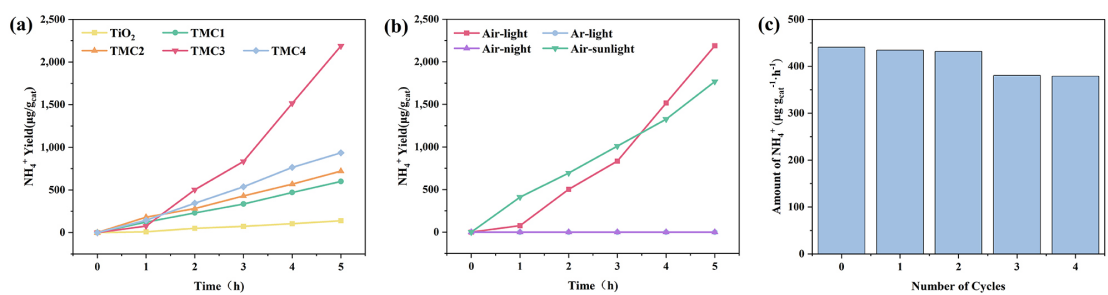

Figure 7. (a) Quantitative determination of the generated NH$_3$ under UV light; (b)Photocatalytic nitrogen fixation tests for TMC3 under different conditions; (c) Catalyst cycling tests for TMC3.

### 3.3.2 Photocatalytic nitrogen fixation mechanism

In order to obtain the conduction band potential of TiO$_2$ and MC, the Mott-Schottky plots of the sample TiO$_2$ and MC were obtained through the electrochemical test. As shown in Figure S5a, b, the 1/C$^2$ to potential curves are obliquely cut to the X axis, and the intersection points are the flat band potentials (E$_{fb}$) of sample TiO$_2$ and MC, which are -0.60 eV and -0.45 eV, respectively. The conduction band potential (E$_{cb}$) is estimated by subtracting 0.1 eV from E$_{fb}$, so the E$_{cb}$ of sample TiO$_2$ is -0.70 eV, which is more negative than that of sample MC (E$_{cb}$= -0.55 eV). The forbidden band width (Eg) of TiO$_2$ is 3.02 eV which is calculated by solid ultraviolet diffuse reflectance spectrum (Figure 4.a). The valence band potential of TiO$_2$ is calculated to be 2.32 eV by the formula $E_{vb} = E_g + E_{cb}$. It has been reported that the reduction potential when $N_2$ is converted to $NH_3$ is -0.092V *vs.* NHE. The oxidation potential of $H_2O$ to $O_2$ is +1.23V *vs.* NHE.[51 52]

Based on the above analysis and previous studies [53], a hypothetical mechanism for photocatalytic nitrogen reduction electron transfer was proposed. In the heterojunction formed by $TiO_2$ and MC, the conduction band potential of $Mo_2C$ satisfies the conversion of $N_2$ to $NH_3$ and is lower than the conduction band potential of $TiO_2$, so that photoelectrons can be transferred from $TiO_2$ to $Mo_2C$. The sample MC prepared in this study has a highly graphitized carbon layer structure (shown in Figure 3.a), and the photoelectrons are easily concentrated on Mo2C due to the excellent electrical conductivity of the carbon layer. Nitrogen is easily adsorbed at the carbon vacancies of the molybdenum atoms to receive photoelectrons, and the activated nitrogen is combined with the protons in water and finally reduced to ammonia. Hydroxyl radicals generated by photocatalytic water splitting of the catalyst were detected in the ESR free radical test (Figure S2.b), confirming the water splitting process. It is inferred from the hydroxyl radical test that the water molecules adsorbed on the surface of $TiO_2$ are oxidized by $TiO_2$ holes and eventually form $O_2$. The following equations related to the reaction are summarized in Scheme 1.

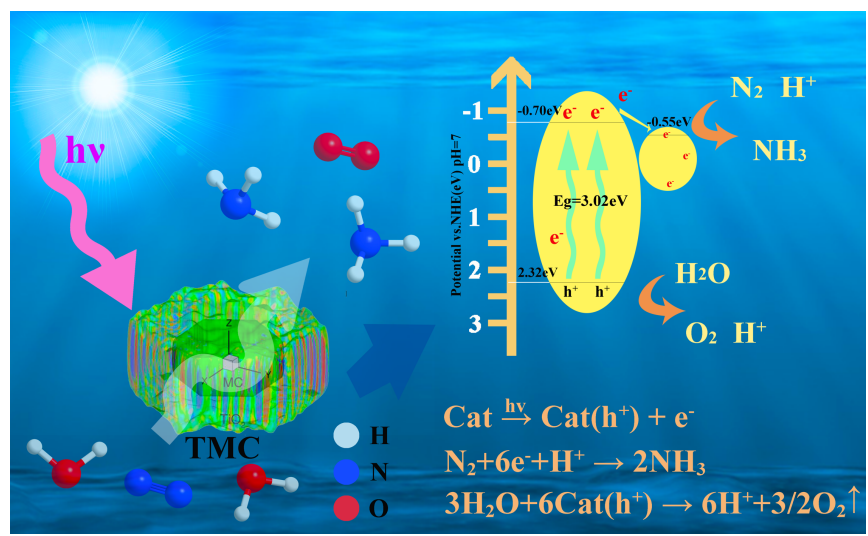

Scheme 1. Schematic illustration showing the mechanism for photocatalytic nitrogen fixation.

## 3.4 Interactions between light and TMCx photocatalytic particle

Numerical simulations via the field-only surface integral method [54-58] were also carried out to study the interactions between light and the synthesized TMCx photocatalytic particles to further confirm that TMCx can be an effective and efficient material for photocatalyst. The focus of the calculation is the absorption cross section of a TMCx photocatalytic particle which represents how much energy from the sunlight can be utilized by such a particle.

Figure 2c and 2d indicate that the TMCx particle should be with a core-shell structure as MC@TiO$_2$ in which the MC core is a thin regular hexagon and the TiO$_2$ is a hexagon with some surface roughness. The typical edge length of the hexagonal MC particle is 2 μm and the thickness was set as 120 nm. The aspect ratio of the TiO$_2$ shell is the same as the MC core, and for the TMC3 particle, the edge length of the TiO$_2$ shell is 1.8 times of the MC core given the densities of and mass ratio between TiO$_2$ and MC. The irregular surface roughness on TiO$_2$ was represented by the Gaussian random particle.

The sunlight is represented by a spectrum of plane electromagnetic waves with time dependence of $\exp(-i\omega t)$ where $\omega$ is the angular frequency of the light and $i$ is the imaginary unit. We screened the wavelength spectrum of sunlight $\lambda$ from 300 nm to 800 nm with a step of 20 nm. The illumination plane wave propagates perpendicular to the thin TMC particle (along $z$-direction) as $\boldsymbol{k}^{\mathrm{inc}} = (0,0,k_0)$ with $k_0 = 2\pi/\lambda$ being the wavenumber, and its electric field polarizes along $x$-direction as $\boldsymbol{E}^{\mathrm{inc}} = \boldsymbol{E}_0 \exp(ik_0 z)$ with $\boldsymbol{E}_0 = (E_0, 0, 0)$ (see Figure 8a). The scattered electromagnetic fields in the solution, $\boldsymbol{E}^{\mathrm{sca}}$ and $\boldsymbol{H}^{\mathrm{sca}}$, and the transmitted fields within the core-shell particle were calculated following the idea in the literature.[57,59] The refractive index of the solution (water) was set as $n_{\mathrm{water}} = 1.33$. The refractive index $n_{\mathrm{MC}}$ and extinction coefficient $k_{MC}$ of MC were calculated by using the first-principles simulations,[60] and the refractive index $n_{TiO_2}$ and extinction coefficient $k_{TiO_2}$ of the traditional photocatalyst material TiO$_2$ can be found in the literature.[61]

Figure 8a displays the distribution of the surface charge density, $q = \boldsymbol{E} \cdot \boldsymbol{n}$, on both surfaces of the MC@ core-shell particle when $\lambda = 540$ nm. Figure 8b presents the variation of absorption cross section $\sigma^{abs} = \int_S \frac{1}{2} \text{Real}[\boldsymbol{E}^{tot} \times (\boldsymbol{H}^{tot})^*] \, d\boldsymbol{S}$ along the wavelengths of sunlight where $\boldsymbol{E}^{tot} = \boldsymbol{E}^{inc} + \boldsymbol{E}^{sca}$ and $\boldsymbol{H}^{tot} = \boldsymbol{H}^{inc} + \boldsymbol{H}^{sca}$, superscript * indicates the conjugate of the field values, and $\boldsymbol{S}$ is a surface that encloses the particle. Obvious absorption of sunlight indicates that TMC is an effective photocatalyst material to harvest solar energy. Also, the trend of $\sigma^{abs}$ across the sunlight spectrum from the simulation in Figure 8b is in good agreement relative to the absorption measured in the experiments as shown in Figure 4a, which means that the assumed core-shell structure of the TMC particle in the simulations is valid.

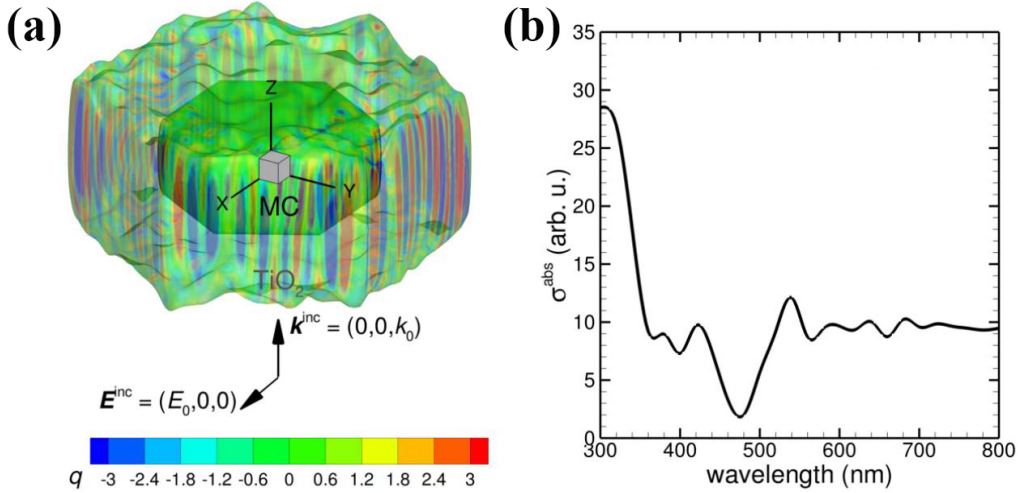

Figure 8. (a) Sketch and surface charge distribution, $q$, of a TMC particle with MC@TiO$_2$ core-shell structure under the illumination of sunlight (Note the length scale in the $z$-direction is 13 times of that in the $xy$-plane for good visualization). (b) Absorption cross sections of a TMC particle across the spectrum of sunlight.

## 4. Conclusion

In summary, a carbon-coated hexagonal Ti/Mo photocatalyst for nitrogen fixation from air directly has been developed. Nitrogen is converted to ammonia under 300W xenon lamp at room temperature and normal pressure, and the reliability of $N_2$ reduction to ammonia has been proved by $^{15}N_2$ isotope labelling. The highest $NH_3$ generation rate of sample TMC3 is achieved to be 432 $\mu g \cdot g_{cat}^{-1} \cdot h^{-1}$, which is 16 times higher than that of pure $TiO_2$. The field-only surface integral method were performed to confirm that it can be an efficient photocatalyst. This study provides a pathway for the feasibility of utilizing $Mo_2C$ in the combination with traditional photocatalyst $TiO_2$ for photocatalytic nitrogen fixation.

## Acknowledgement

We are thankful to the National Natural Science Foundation of China (21978061) and Zhejiang Key Laboratory of Green Pesticides and Cleaner Production Technology for providing financial support. Q. Sun acknowledges the support by the Australian Research Council (ARC) through grants DE150100169, FT160100357 and CE140100003. This research was undertaken with the assistance of resources from the National Computational Infrastructure (NCI Australia), an NCRIS enabled capability supported by the Australian Government.